\documentclass{elsarticle} 
\usepackage{lineno,hyperref}
\usepackage{amsmath}
\usepackage[english]{babel}
\usepackage{graphicx}

\makeatletter
\def\ps@pprintTitle{%
 \let\@oddhead\@empty
 \let\@evenhead\@empty
 \def\@oddfoot{}%
 \let\@evenfoot\@oddfoot}
\makeatother

\modulolinenumbers[5]

\begin{document}
\begin{frontmatter}
\title{Self-consistent description of interacting phonons in a crystal lattice}

\author[kipt]{Yuri Poluektov}
\ead{yuripoluektov@kipt.kharkov.ua}
\author[kipt]{Volodymyr Savchenko\corref{corresponding_author}}
\cortext[corresponding_author]{Corresponding author}
\ead{vovasvch@gmail.com}

\address[kipt]{Akhiezer Institute for Theoretical Physics, NSC "Kharkov Institute of Physics and Technology", 1, Akademicheskaya Str., Kharkov UA-61108, Ukraine}

\begin{keyword}
 Phonon \sep specific heat \sep phonon-phonon interaction \sep Debye energy \sep quasiparticle
\end{keyword}

\begin{abstract}
  Self-consistent approach for interacting phonons description in lattice, which generalizes Debye model, is proposed. Notion of "self-consistent" phonons is introduced, speed of which depends on temperature and is determined from non-linear equation. Debye energy is also a function of temperature in this approach. Thermodynamics of "self-consistent" phonon gas is constructed. It is shown, that at low temperatures there is a correction proportional to the seventh power of temperature to the cubic law of specific heat dependence on temperature. This may be one of the reasons why cubic law for specific heat is observed only at rather low temperatures. At high temperatures the theory predicts linear deviation from Dulong--Petit law, which is  observed experimentally.
\end{abstract}
  
\end{frontmatter}
\linenumbers

\section{Introduction}
Quantum mechanics was first used to describe specific heat of solid by Einstein \cite{Einstein}. In his theory solid body was considered as a set of quantum harmonic oscillators. Einstein succeeded in describing specific heat deviation from Dulong-Petit law at low temperatures, but obtained rate of specific heat decrease was exponential and substantially differed from experimentally observed power law. A different approach for quantum description of solid state was proposed by Debye \cite{Debye}. In his model solid state is considered as continuous medium, excitations in which are quantized. Debye obtained correct proportional to the third power of temperature behavior of low temperature specific heat and introduced fundamental notion of Debye temperature (energy or frequency) into solid state physics. In fact, as it is clear now, Einstein and Debye models describe solid state from different points of view and supplement each other. Einstein model describes single--particle excitations and preserves its significance e.g. for description of optical branches of oscillations, while Debye model gives a quantum description of collective excitations of solid. Debye theory was so successful in it's simplicity and obviousness, that it is still applied more often than more detailed Born--von Karman theory of the crystal lattice, which was also published at that time \cite{Born_Karman}.\\
\indent In Debye model excitations of medium are described as a gas of non-interacting at any temperature quasiparticles -- phonons. However it is easy to see, that number of phonons increases with increase of temperature, and thus interaction between them becomes more and more significant. In the present work Debye model is generalized on the case of phonon gas, interaction in which is taken into account in self-consistent field approximation. Accounting for phonon--phonon interaction leads to a conclusion, that speed of such "self-consistent" phonons becomes a function of temperature and is not considered as given, but is obtained from solving non-linear algebraic equation, which in turn is derived from the free energy minimum condition. "Self-consistent" Debye energy in such approach is also a function of temperature. Thermodynamic quantities of a gas of such phonons are calculated and, in particular, phonon specific heat. It is shown, that at low temperatures there is a correction to the cubic law of specific heat dependence on temperature, which is proportional to the seventh power of temperature. This, apparently, may be one of the reasons why the cubic law for specific heat is only observed at rather low temperatures. At high temperatures theory predicts linear on temperature deviation of specific heat behavior from Dulong-Petit law, which is observed experimentally.
\section{Model definition}
Phonons in crystal lattice are described by a Hamilton operator density
\begin{equation}\label{initial_hamiltonian}
H(\mathbf{r}) = \frac {\pi_{a}^{2}(\mathbf{r})} {2 \rho} + U_{2}(\mathbf{r}) + U_{3}(\mathbf{r}) + U_{4}(\mathbf{r}),
\end{equation}
where quadratic, cubic and forth power terms on deformation tensor
\begin{equation}\label{deformation_tensor_def}
u_{ij} = \frac {1} {2} (\nabla_{j} u_{i} + \nabla_{i} u_{j} + \nabla_{i} u_{a}\nabla_{j} u_{a})
\end{equation}
have the form
\begin{equation}\label{u2_def}
U_{2} = \frac {1} {2} \lambda_{aibj} u_{ai} u_{bj},
\end{equation}
\begin{equation}\label{u3_def}
U_{3} = \frac {1} {6} \lambda_{aibjck} u_{ai} u_{bj} u_{ck},
\end{equation}
\begin{equation}\label{u4_def}
U_{4} = \frac {1} {24} \lambda_{aibjckdl} u_{ai} u_{bj} u_{ck} u_{dl},
\end{equation}
$\pi_{a}(\mathbf{r})$ is canonical momentum, $\rho$ is density. As a result of deformation tensor symmetry $u_{ij} = u_{ji}$ elastic moduli are symmetrical with respect to both permutation of index pairs and permutation of indices inside pairs, in particular
\begin{equation}\label{lambda_symmetry}
\lambda_{aibj} = \lambda_{bjai} = \lambda_{jbai} = \lambda_{jbia}.
\end{equation}
Analogous symmetry rules are applied to elastic moduli of third and forth orders.\\
\indent Deformation tensor  (\ref{deformation_tensor_def}) contains quadratic terms on the gradient of deformation vector $\mathbf{u} (\mathbf{r})$ as well as linear ones. Therefore, potential energies (\ref{u2_def},\ref{u3_def}) accurate within fourth order of deformation vector gradients can be written in the form $U_{2} = U_{2}^{(2)} + U_{2}^{(3)} + U_{2}^{(4)}$, $U_{3} = U_{3}^{(3)} + U_{3}^{(4)}$, where
\begin{equation}\label{u2_grad_expand2}
U_{2}^{(2)} = \frac {1} {2} \lambda_{aibj} \nabla_{i} u_{a} \nabla_{j} u_{b},
\end{equation}
\begin{equation}\label{u2_grad_expand3}
U_{2}^{(3)} = \frac {1} {2} \lambda_{aibj} \nabla_{i} u_{a} \nabla_{j} u_{c} \nabla_{b} u_{c},
\end{equation}
\begin{equation}\label{u2_grad_expand4}
U_{2}^{(4)} = \frac {1} {8} \lambda_{aibj} \nabla_{a} u_{c} \nabla_{i} u_{c} \nabla_{b} u_{s} \nabla_{j} u_{s}, 
\end{equation}
\begin{equation}\label{u3_grad_expand3}
U_{3}^{(3)} = \frac {1} {6} \lambda_{aibjck} \nabla_{i} u_{a} \nabla_{j} u_{b} \nabla_{k} u_{c},
\end{equation}
\begin{equation}\label{u3_grad_expand4}
U_{3}^{(4)} = \frac {1} {4} \lambda_{aibjck} \nabla_{i} u_{a} \nabla_{j} u_{b} \nabla_{k} u_{s} \nabla_{c} u_{s}.
\end{equation}
Thus, accurate within the fourth order of deformation vector gradients Hamiltonian density takes the form
\begin{equation}\label{hamiltonian_fourth}
H(\mathbf{r}) = \frac {\pi_{a}^{2} (\mathbf{r})} {2 \rho} + \frac{1} {2} \lambda_{aibj} \nabla_{i} u_{a} \nabla_{j} u_{b} + \tilde U_{3} + \tilde U_{4},
\end{equation}
where
\begin{equation}\label{u3_tilde}
\tilde U_{3} = \frac {1} {2} \lambda_{aibj} \nabla_{i} u_{a} \nabla_{j} u_{c} \nabla_{b} u_{c} + \frac {1} {6} \lambda_{aibjck} \nabla_{i} u_{a} \nabla_{j} u_{b} \nabla_{k} u_{c},
\end{equation}
\begin{equation}\label{u4_tilde}
\begin{split}
\tilde U_{4} = \frac {1} {8} \lambda_{aibj} \nabla_{a} u_{c} \nabla_{i} u_{c} \nabla_{b} u_{s} \nabla_{j} u_{s} + \\
+ \frac {1} {4} \lambda_{aibjck} \nabla_{i} u_{a} \nabla_{j} u_{b} \nabla_{k} u_{s} \nabla_{c} u_{s} + \\
+ \frac {1} {24} \lambda_{aibjckdl} \nabla_{i} u_{a} \nabla_{j} u_{b} \nabla_{k} u_{c} \nabla_{l} u_{d}.
\end{split}
\end{equation}
Under quantum description deformation vector $u_{a}(\mathbf{r}) = u_{a}^{\dagger}(\mathbf{r})$ and canonical momentum $\pi_{a}(\mathbf{r})$ should be considered as operators, for which the following commutation relations are true
\begin{equation}\label{comm_pi_u}
\pi_{a}(\mathbf{r}) u_{b}(\mathbf{r'}) - u_{b}(\mathbf{r'}) \pi_{a}(\mathbf{r}) = - i \hbar \delta_{ab} \delta(\mathbf{r} - \mathbf{r'}),
\end{equation}
\begin{equation}\label{comm_u_u}
u_{a}(\mathbf{r}) u_{b}(\mathbf{r'}) - u_{b}(\mathbf{r'}) u_{a}(\mathbf{r}) = 0,
\end{equation}
\begin{equation}\label{comm_pi_pi}
\pi_{a}(\mathbf{r}) \pi_{b}(\mathbf{r'}) - \pi_{b}(\mathbf{r'}) \pi_{a}(\mathbf{r}) = 0.
\end{equation}
Elastic moduli, generally speaking, can be functions of temperature, but in the present work, as in the Debye model, we will neglect such dependencies in the following and only will take into account temperature dependencies of observed quantities connected with phonon excitations. Full Hamiltonian $H = \int H(\mathbf{r})d \mathbf{r}$ is a sum of free phonons Hamiltonian and interaction Hamiltonian $H = H_{0} + H_{I}$, where
\begin{equation}\label{hamiltonian_free_phonons}
H_{0} = \int \left \{ \frac {\pi_{a}^{2} (\mathbf{r})} {2 \rho} + \frac{1} {2} \lambda_{aibj} \nabla_{i} u_{a} \nabla_{j} u_{b} \right \} d \mathbf{r},
\end{equation}
\begin{equation}\label{hamiltonian_interaction}
H_{I} = \int \left [ \tilde U_{3}(\mathbf{r}) + \tilde U_{4}(\mathbf{r}) \right ] d \mathbf{r}.
\end{equation}
Before proceeding to detailed construction of the model let us introduce some functions, which will be useful later. Generalized Debye functions are defined as follows
\begin{equation}\label{generalized_debye}
D_{n}(x) = \frac {n} {x^{n}} \int\limits_{0}^{x} \frac {z^{n} dz} {e^{z} -1}, \quad (n \ge 1).
\end{equation}
Actually only these functions at $n = 1,2,3$ will be necessary. Standard Debye function in these notations is $D_{3}(x)$ \cite{Landau_Lifshitz}. Functions (\ref{generalized_debye}) can be presented in the form
\begin{equation}\label{exp_form}
D_{n} = \frac {n} {x^{n}} \left [ n! \zeta(n+1) - \sum_{m=0}^{\infty} \int\limits_{x}^{\infty} e^{-(m+1)z} z^{n} dz \right ],
\end{equation}
so that at $x \gg 1$ neglecting exponentially small terms one obtains
\begin{equation}\label{debye_big_x}
D_{1}(x) \approx \frac {\pi^{2}} {6x}, \quad D_{2}(x) \approx \frac {4} {x^{2}} \zeta(3), \quad D_{3}(x) \approx \frac {\pi^{4}} {5x^{3}}.
\end{equation}
$\zeta(3) \approx 1.202$ is Riemann zeta function. At $x \ll 1$:
\begin{equation}\label{debye_small_x}
D_{n}(x) \approx 1 - \frac {n} {2(n + 1)} x + \frac {n} {12(n + 2)} x^{2}.
\end{equation}
Besides that function $\Phi(x)$ will be used in the following considerations
\begin{equation}\label{Phi_def}
\Phi (x) \equiv 1 + \frac {8} {3x} D_{3}(x),
\end{equation}
\begin{equation}\label{Phi_approx}
\Phi (x) \approx \frac {8} {3x} + \frac{2} {15} x, \; (x \ll 1), \quad \Phi(x) \approx 1 + \frac {8 \pi^{4}} {15 x^{4}}, \; (x \gg 1).
\end{equation}
\section{Self-consistent description of interacting phonon system}
\indent Let us consider an interacting phonon system using self-consistent field approach, developed in \cite{Poluektov1} for fermionic and in \cite{Poluektov2,Poluektov3} for bosonic systems. Implementation of such method on the example of anharmonic oscillator is demonstrated in \cite{Poluektov4}.\\
\indent Let us divide Hamiltonian into a sum of two terms
\begin{equation}\label{division}
H = H_{S} + H_{C},
\end{equation}
where renormalized Hamiltonian describing "free" phonons has the form
\begin{equation}\label{H_s}
H_{S} = \int \left [ \frac {\pi_{a}^{2}} {2 \rho} + \frac {\tilde \lambda} {2} \nabla_{i} u_{a} \nabla_{i} u_{a} \right ] d \mathbf{r} + \varepsilon_{0},
\end{equation}
and correlation Hamiltonian
\begin{equation}\label{H_c}
H_{C} = \int \left [ \frac {1} {2}(\lambda_{aibj} - \tilde \lambda \delta_{ij} \delta_{ab}) \nabla_{i} u_{a} \nabla_{j} u_{b} + \tilde U_{3} + \tilde U_{4} \right ] d \mathbf{r} - \varepsilon_{0}
\end{equation}
describes interaction of such phonons. Self-consistent approximation Hamiltonian (\ref{H_s}) contains only one effective elastic modulus $\tilde \lambda$ and describes the phonon system in isotropic approximation, when phonons with arbitrary polarization have the same speed. Besides that, $H_{S}$ includes a non-operator term $\varepsilon_{0}$, accounting for which is substantial. Thus, by means of elastic modulus renormalization, interaction between initial phonons is taken into account in isotropic approximation in Hamiltonian (\ref{H_s}), and Hamiltonian (\ref{H_c}) takes into account residual interaction not included in self-consistent field model. Expansions of field operators are
\begin{equation}\label{pi_expansion}
\pi_{a}(\mathbf{r}) = - \frac {1} {\sqrt{V}} \sum_{\mathbf{k}, \alpha} \sqrt{\frac {\rho \hbar \omega(\mathbf{k}, \alpha)} {2}} e_{a} (\mathbf{k}, \alpha) \chi_{\mathbf{k} \alpha} e^{i \mathbf{kr}},
\end{equation}
\begin{equation}\label{u_expansion}
u_{a}(\mathbf{r}) = \frac {1} {\sqrt{V}} \sum_{\mathbf{k}, \alpha} \sqrt{\frac {\hbar} {2 \rho \omega(\mathbf{k}, \alpha)}} e_{a} (\mathbf{k}, \alpha) \psi_{\mathbf{k} \alpha} e^{i \mathbf{kr}},
\end{equation}
where $\mathbf{e}(\mathbf{k}, \alpha)$ are polarization vectors which satisfy the following relations
\begin{equation}\label{e_orth}
\begin{split}
\mathbf{e}(\mathbf{k}, \alpha) \mathbf{e}^{*}(\mathbf{k}, \alpha') = \delta_{\alpha \alpha'}, \quad \sum_{\alpha} e_{i}^{*}(\mathbf{k}, \alpha) e_{j}(\mathbf{k}, \alpha) = \delta_{ij},\\
\mathbf{e}(- \mathbf{k}, \alpha) = \mathbf{e}^{*}(\mathbf{k}, \alpha).
\end{split}
\end{equation}
In (\ref{pi_expansion},\ref{u_expansion}) operators
\begin{equation}\label{chi_psi}
\begin{split}
\psi_{\mathbf{k} \alpha} = \psi_{-\mathbf{k} \alpha}^{\dagger} = b_{\mathbf{k} \alpha} + b_{-\mathbf{k} \alpha}^{\dagger}, \\
\chi_{\mathbf{k} \alpha} = \chi_{-\mathbf{k} \alpha}^{\dagger} = i \left(b_{\mathbf{k} \alpha} - b_{-\mathbf{k} \alpha}^{\dagger} \right)
\end{split}
\end{equation}
are introduced. Lattice is considered to be simple, so that $\alpha = 1,2,3$. Phonon creation $b_{\mathbf{k} \alpha}^{\dagger}$ and annihilation $b_{\mathbf{k} \alpha}$ operators satisfy usual commutation laws $[b_{\mathbf{k}\alpha},b_{\mathbf{k'} \alpha'}^{\dagger}] = \delta_{\mathbf{k}\mathbf{k'}} \delta_{\alpha \alpha'}$, $[b_{\mathbf{k} \alpha},b_{\mathbf{k'} \alpha'}] = [b_{\mathbf{k} \alpha}^{\dagger},b_{\mathbf{k'} \alpha'}^{\dagger}] = 0 $.\\
\indent Self-consistent Hamiltonian (\ref{H_s}) in terms of creation and annihilation operators acquires the form 
\begin{equation}\label{H_annihilation}
H_{S} = \hbar \sum_{\mathbf{k}, \alpha} \omega (k) b_{\mathbf{k}, \alpha}^{\dagger} b_{\mathbf{k}, \alpha} + \frac {3} {2} \hbar \sum_{\mathbf{k}} \omega(k) + \varepsilon_{0},
\end{equation}
where $\omega(k) = c_{S} k$ and speed of phonons with arbitrary polarization is $c_{S} = \sqrt{\tilde \lambda / \rho}$. It should be noted, that neglecting phonon--phonon interaction in Hamiltonian (\ref{H_annihilation}) leads to Debye theory.\\
\indent In what follows the averaging is made using statistical operator
\begin{equation}\label{stat_op_def}
\hat \rho = \exp \beta(F - H_{S}),
\end{equation}
where $\beta = 1/T$ is inverse temperature. Normalizing condition $Sp \hat \rho = 1$ leads to expression for free energy
\begin{equation}\label{free_en}
F = \varepsilon_{0} + \frac {3 \hbar} {2} \sum_{\mathbf{k}} \omega(k) + 3T \sum_{\mathbf{k}} \ln \left(1 - e^{-\beta \hbar \omega(k)} \right).
\end{equation}
Value of $\varepsilon_{0}$ is obtained from usual for mean field theory condition $\langle H \rangle = \langle H_{S} \rangle$ \cite{Poluektov1}, so that
\begin{equation}\label{epsilon_eq}
\begin{split}
\varepsilon_{0}=\int \left[ \frac{1}{2}\left( \lambda_{aibj} - \tilde\lambda \delta_{ij}\delta_{ab} \right)\left\langle \nabla_{i} u_{a}\nabla_{j} u_{b} \right\rangle + \left\langle \tilde U_{3} \right\rangle +\left\langle \tilde U_{4} \right\rangle  \right] d \mathbf{r}.
\end{split}
\end{equation}
Here 
\begin{equation}\label{mean2}
\left\langle \nabla_{i} u_{a} \nabla_{j} u_{b} \right\rangle =\frac{\hbar }{2\rho V} \delta_{ab} \sum_{\mathbf{k}} \frac{k_{i} k_{j}} {\omega (k)} \left(1+2 f_{k}\right), \quad \left\langle \tilde U_{3} \right\rangle = 0,
\end{equation}
\begin{equation}\label{mean4}
\begin{split}
\left\langle \nabla_{i} u_{a} \nabla_{j} u_{b} \nabla_{k} u_{c} \nabla_{l} u_{d} \right\rangle = \frac{1}{V^{2}} \left( \frac{\hbar } {2 \rho} \right)^{2} \sum_{\mathbf{k}_{1},\mathbf{k}_{2}} \frac{ ( 1 + 2f_{k_{1}}) (1 + 2 f_{k_{2}})} {\omega (k_{1}) \omega (k_{2})} \times \\
\times \left[ k_{1i} k_{1j} k_{2k} k_{2l} \delta_{ab} \delta_{cd}  + k_{1i} k_{1k} k_{2j} k_{2l} \delta_{ac} \delta_{bd} + k_{1i} k_{1l} k_{2j} k_{2k} \delta_{ad} \delta_{bc} \right],
\end{split}
\end{equation}
\begin{equation}\label{mean4_sum1}
\begin{split}
\left\langle \nabla_{i} u_{a} \nabla_{j} u_{b} \nabla_{k} u_{s} \nabla_{c} u_{s} \right\rangle = \frac{1}{V^{2}} \left( \frac{\hbar } {2 \rho} \right)^{2} \delta_{ab} \sum_{\mathbf{k}_{1},\mathbf{k}_{2}} \frac{ ( 1 + 2f_{k_{1}}) (1 + 2 f_{k_{2}})} {\omega (k_{1}) \omega (k_{2})} \times \\
\times \left[ 3 k_{1i} k_{1j} k_{2k} k_{2c} + k_{1i} k_{1k} k_{2j} k_{2c} + k_{1i} k_{1c} k_{2j} k_{2k} \right],
\end{split}
\end{equation}
\begin{equation}\label{mean4_sum2}
\begin{split}
\left\langle \nabla_{i} u_{c} \nabla_{a} u_{c} \nabla_{b} u_{s} \nabla_{j} u_{s} \right\rangle = \frac{3}{V^{2}} \left( \frac{\hbar } {2 \rho} \right)^{2} \sum_{\mathbf{k}_{1},\mathbf{k}_{2}} \frac{ ( 1 + 2f_{k_{1}}) (1 + 2 f_{k_{2}})} {\omega (k_{1}) \omega (k_{2})} \times \\
\times \left[ 3k_{1i} k_{1a} k_{2b} k_{2j} + k_{1i} k_{1b} k_{2a} k_{2j} + k_{1i} k_{1j} k_{2a} k_{2b} \right],
\end{split}
\end{equation}
In (\ref{mean2},\ref{mean4},\ref{mean4_sum1},\ref{mean4_sum2})
\begin{equation}\label{distrib_func}
f_{k} = \frac {1} {e^{\beta \hbar \omega(k)} - 1}
\end{equation}
is phonon distribution function. From these equations follows
\begin{equation}\label{epsilon0_sum}
\begin{split}
\varepsilon_{0} = \frac {3 \hbar} {2 c_{S}} \left[ \frac {1} {3 \rho}\sum_{\mathbf{k}} \frac{k_{i} k_{j} \lambda_{aiaj}} {k} \left( f_{k} +\frac {1} {2} \right) - c_{S}^{2} \sum_{\mathbf{k}} k\left(f_{k} + \frac{1} {2} \right) \right] + \frac {\hbar^{2}} {8V \rho^{2} c_{S}^{2}} I,
\end{split}
\end{equation}
where
\begin{equation}\label{I_def}
\begin{split}
I \equiv \sum_{\mathbf{k}_{1}, \mathbf{k}_{2}} \frac{\left( f_{k_{1}} +\frac{1} {2} \right) \left( f_{k_{2}} +\frac {1} {2} \right)}{ k_{1} k_{2}}\left\{ \lambda_{aiajbkbl} k_{1i} k_{1j} k_{2k} k_{2l} + \right. \\ 
 \left. + 2\lambda_{aiajck} \left[ 3k_{1i} k_{1j} k_{2k} k_{2c} + 2k_{1i} k_{1k} k_{2j} k_{2c} \right] + \right. \\ 
 \left. + 3 \lambda_{aibj} \left[ 3 k_{1i} k_{1a} k_{2b} k_{2j} + 2 k_{1i} k_{1b} k_{2a} k_{2j} \right] \right\}.
\end{split}
\end{equation}
Renormalized in consequence of interaction phonon speed $c_{S}$ is obtained from the free energy (\ref{free_en}) minimum condition $\partial F / \partial c_{S} = 0$. As a result one obtains non-linear equation for speed of "new" phonons
\begin{equation}\label{c_s_2}
c_{S}^{2} = \frac {2 \pi^{2}} {3\rho V} J^{-1} \sum_{\mathbf{k}} \frac{ k_{i} k_{j} \lambda_{aiaj}} {k} \left( f_{k} + \frac {1} {2} \right) +\frac{\hbar \pi^{2}} {6 \rho^{2} V^{2} c_{S}} \frac{I}{J},
\end{equation}
where
\begin{equation}\label{j_def}
J=\int\limits_{0}^{ k_{D}} \left( f_{k} +\frac{1}{2} \right) k^{3} dk.
\end{equation}
Upper limit of integration in (\ref{j_def}) is Debye wave number $k_{D}$. It is obtained from condition, that sphere of $k_{D}$ radius in $k$--space should have a number of points, which coincides with the number of particles in the system \cite{Zaiman}
\begin{equation}\label{k_d}
\frac{V 4\pi }{ \left( 2\pi  \right)^{3}} \int\limits_{0}^{k_{D}} k^{2} dk = \frac{V k_{D}^{3}}{6 \pi^{2}} = N, \quad k_{D} = \left( 6 \pi^{2} N/V \right)^{1/3}.
\end{equation}
Equation (\ref{c_s_2}) is true for crystals of arbitrary symmetry.\\
\indent In the absence of interaction and neglecting non-linear effects, phonons in this theory are the same as in Debye theory. It is natural to name them "bare" or "Debye" phonons. Phonons, speed of which is renormalized according to (\ref{c_s_2}), as a consequence of interaction, we will call "self-consistent". Even neglecting temperature dependencies of $\lambda_{aiaj}$ moduli, as it is supposed in Debye theory, renormalized speed $c_{S}$ in this approach significantly depends on temperature, because it is expressed through integrals of distribution function (\ref{distrib_func}).\\
\indent It is possible to ascertain, that under fulfillment of the conditions (\ref{epsilon_eq}) and (\ref{c_s_2}) equation $\partial \Gamma / \partial c_{S} = 0$ is true, where $\Gamma = \left \langle (H - H_{S}) \right \rangle$ is an average of a difference between exact and self-consistent Hamiltonians. Thus, it turns out that for introduced quadratic Hamiltonian (\ref{H_s}) with conditions (\ref{epsilon_eq}) and (\ref{c_s_2}) satisfied, the quantity of $\Gamma$ becomes minimal. Consequently, parameter $\tilde \lambda$ is chosen in a such way, that approximation Hamiltonian (\ref{H_s}) turns out to be the closest of all quadratic Hamiltonians to the exact Hamiltonian (\ref{division}).
\section{Self-consistent phonons in isotropic medium}
\indent Let us consider interacting phonons in isotropic medium in more detail. In this case the quadratic elastic moduli tensor has the form
\begin{equation}\label{lambda_4_iso}
\lambda_{aibj} = \lambda \delta_{ai} \delta_{bj} + \mu (ij,ab),
\end{equation}
where $\lambda, \mu$ are Lam\'e coefficients. Notation 
\begin{equation}\label{bracket_def}
(ij,ab) \equiv \delta_{ij} \delta_{ab} + \delta_{ia} \delta_{jb}
\end{equation}
is introduced for brevity, which satisfies symmetry conditions
\begin{equation}\label{bracket_symmetry}
(ij,ab) = (ji,ba) = (ab,ij) = (ba,ji).
\end{equation}
Anharmonic elastic moduli tensors in isotropic elastic medium have the forms
\begin{equation}\label{lambda_6_iso}
\begin{split}
\lambda_{aibjck} = B_{1} \delta_{ai} \delta_{bj} \delta_{ck} + B_{2}\left[ \delta_{ai} (jk,cb) + \delta_{bj} (ik,ca) + \delta_{ck} (ij,ba) \right] + \\ 
+ B_{3} \left[ \delta_{ac} (ij,bk) + \delta_{ak} (ij,bc) + \delta_{ic} (jk,ab) + \delta_{ik} (ab,jc) \right],
\end{split}
\end{equation}
\begin{equation}\label{lambda_8_iso}
\begin{split}
\lambda_{aibjckdl} = C_{1} \delta_{ai} \delta_{bj} \delta_{ck} \delta_{dl} + C_{2} \lambda_{aibjckdl}^{(2)} + \\
+ C_{3} \lambda_{aibjckdl}^{(3)} + C_{4} \lambda_{aibjckdl}^{(4)} + C_{5} \lambda_{aibjckdl}^{(5)} \;,
\end{split}
\end{equation}
where
\begin{equation}\label{lambda_8_2_iso}
\begin{split}
\lambda_{aibjckdl}^{(2)} \equiv \delta_{ai} \delta_{bj} (lk,cd) + \delta_{ai} \delta_{ck} (jl,db) + \\
+ \delta_{ai} \delta_{dl} (jk,cb) + \delta_{bj} \delta_{ck} (il,da) + \\
+ \delta_{bj} \delta_{dl} (ik,ca) + \delta_{ck} \delta_{dl} (ij,ba),
\end{split}
\end{equation}
\begin{equation}\label{lambda_8_3_iso}
\begin{split}
\lambda_{aibjckdl}^{(3)} \equiv \delta_{ai} \left[ \delta_{bd} (jk,cl) + \delta_{bl} (jk,cd) + \delta_{jd} (kl,bc) + \delta_{jl} (bc,kd) \right] + \\ 
 + \delta_{bj} \left[ \delta_{ac} (kl,di) + \delta_{ci} (kl,da) +  \delta_{ka} (li,cd) + \delta_{ik} (ad,lc) \right] + \\ 
 + \delta_{ck} \left[ \delta_{bd} (il,ja) + \delta_{dj} (il,ba) +  \delta_{lb} (ij,da) + \delta_{jl} (ab,di) \right] + \\ 
 + \delta_{dl} \left[ \delta_{ac} (ij,bk) + \delta_{ak} (ij,bc) +  \delta_{ic} (jk,ab) + \delta_{ik} (ab,jc) \right],
\end{split}
\end{equation}
\begin{equation}\label{lambda_8_4_iso}
\lambda_{aibjckdl}^{(4)} \equiv (il,da) (jk,cb) + (ik,ca)(jl,db) + (ij,ba) (kl,dc),
\end{equation}
\begin{equation}\label{lambda_8_5_iso}
\begin{split}
\lambda_{aibjckdl}^{(5)} \equiv \\
  (ab,ci) (jl,dk) + (ij,ca) (kl,db) + (ik,ba) (jl,dc) + (ij,ka) (bl,dc) + \\
 + (ab,jc) (il,dk) + (ij,bc) (kl,da) + (ab,jk) (il,dc) + (ij,bk) (cl,da) + \\
 + (ad,ci) (jk,lb) + (jl,cb) (ik,da) + (jk,db) (il,ca) + (il,ka) (cj,bd).
\end{split}
\end{equation}
Quantity (\ref{I_def}) can be expressed as
\begin{equation}\label{I_simpl}
I \equiv \sum_{\mathbf{k}_{1},\mathbf{k}_{2}} \frac {\left( f_{k_{1}} + \frac {1} {2} \right) \left( f_{k_{1}} + \frac {1} {2} \right)} {k_{1} k_{2}} \left[ V_{0} k_{1}^{2} k_{2}^{2} + V_{1} \left( \mathbf{k}_{1}\mathbf{k}_{2} \right)^{2} \right]
\end{equation}
where
\begin{equation}\label{V_0}
V_{0} = 9 \lambda + 6 \mu + 6 B_{1} + 32 B_{2} + 32 B_{3} + C_{1} + 8 C_{2} + 8 C_{3} + 18 C_{4} + 28 C_{5}, 
\end{equation}
\begin{equation}\label{V_1}
V_{1} = 6 \lambda + 24 \mu + 4 B_{1} + 48 B_{2} + 88 B_{3} + 8 C_{2} + 40 C_{3} + 10 C_{4} + 68 C_{5}. 
\end{equation}
Taking into consideration these relations and integrating over angles in (\ref{I_simpl}) one obtains the following equation, which determines the self-consistent phonon speed in isotropic elastic medium
\begin{equation}\label{c_s_iso}
c_{S}^{2} = c_{0}^{2} + \frac {\hbar} {24 \pi^{2} \rho^{2} c_{S}} \left( V_{0} + \frac {V_{1}} {3} \right)J,
\end{equation}
where the averaged phonon speed without taking into account interaction $c_{0}$ is defined by formula
\begin{equation}\label{c_0_def}
c_{0}^{2} = \frac {1} {3} (2 c_{t}^{2} + c_{l}^{2}) = \frac {(\lambda + 4 \mu)} {3 \rho},
\end{equation}
and longitudinal and transverse speeds of sound are defined by known equations $c_{l}^{2} = (\lambda + 2 \mu) / \rho$, $c_{t}^{2} = \mu / \rho$. It should be noted, that definition (\ref{c_0_def}) differs from Debye average speed $c_{D}$ definition \cite{Landau_Lifshitz}
\begin{equation}\label{c_d_def}
\frac {1} {c_{D}^{3}} = \rho^{3/2} \left[ \frac {2} {\mu^{3/2}} + \frac {1} {(\lambda + 2\mu)^{3/2}} \right] .
\end{equation}
Authors of the present work consider definition (\ref{c_0_def}) as more natural, however, since average speed enters the definition of Debye energy, which is a phenomenological parameter of the theory, then differences in definitions (\ref{c_0_def}) and (\ref{c_d_def}) does not influence the theory structure. \\
\indent Standard Debye energy in defined as $\Theta_{D} \equiv \hbar c_{0} k_{D}$ \cite{Zaiman}. So far as a new self-consistent phonon speed $c_{S}$ appears besides bare phonon speed $c_{0}$, it is natural to define "self-consistent Debye energy" $\tilde \Theta_{D} \equiv \hbar c_{S} k_{D}$, which, contrary to standard Debye energy $\Theta_{D}$, is a function of temperature. It should be noted, that in spite the fact that in original Debye theory formulation energy $\Theta_{D}$ is supposed to be independent from temperature, practically experimental data is often represented in form of Debye energy as a function of temperature \cite{Reislend}. In the considered approach self-consistent Debye energy significantly depends on temperature as a consequence of phonon--phonon interaction in the framework of the theory itself, and, therefore, the given approach better reflects the real situation. However, drawbacks of Debye theory connected with oversimplified choice of spectral density, which does not take into account the details of lattice structure, persist in the proposed approach. \\
\indent It is interesting to calculate a number of phonons as a function of temperature
\begin{equation}\label{number_phonons}
N_{ph} = 3 \sum_{\mathbf{k}} f_{k} = \frac {9} {2} N \frac {T} {\tilde\Theta_{D}} D_{2} \left( \frac {\tilde\Theta_{D}} {T} \right).
\end{equation}
Taking into consideration properties (\ref{debye_big_x}) and (\ref{debye_small_x}) one finds, that at low temperatures $T \ll \tilde\Theta_{D}$ dependency is cubic $N_{ph} / N \approx 18 \zeta(3) (T / \tilde\Theta_{D})^{3}$, and at high temperatures $T \gg \tilde\Theta_{D}$ dependency is linear $N_{ph} / N \approx (9/2) (T / \tilde\Theta_{D})$. Thus, contribution of interaction between phonons should be more and more significant with the rise of temperature, due to number of phonons increase. At $T \sim \tilde\Theta_{D}$ a number of phonons is of the same order of magnitude as a number of particles. Phonons satisfy Bose--Einstein statistics and therefore there can be any number of phonons in a state with given momentum, so it should not be surprising, that number of phonons can exceed number of particles in solid body. \\
\indent It is convenient to introduce notation $\sigma$ for ratio of renormalized as consequence of phonon--phonon interaction speed $c_{S}$ to initial speed $c_{0}$, or, which is the same, ratio of self-consistent Debye energy to standard one
\begin{equation}\label{sigma_def}
\sigma \equiv \frac {c_{S}} {c_{0}} = \frac {\tilde \Theta_{D}} {\Theta_{D}}.
\end{equation}
For further analysis, it is suitable to express equation (\ref{c_s_iso}) in dimensionless form, taking into consideration introduced notation (\ref{sigma_def})
\begin{equation}\label{main_dimless}
(\sigma^{2} - 1) \sigma = \Lambda \Phi \left(\frac {\sigma} {\tau} \right)
\end{equation}
where $\tau \equiv \frac {T} {\Theta_{D}}$ is dimensionless temperature. In (\ref{main_dimless}) it was taken into account, that $J = \frac {k_{D}^{4}} {8} \Phi \left(\frac {\sigma} {\tau} \right)$, and function $\Phi$ is defined in (\ref{Phi_def}). Only one dimensionless parameter, containing characteristics of a given system, enters into equation (\ref{main_dimless})
\begin{equation}\label{Lambda_def}
\Lambda \equiv \frac {\Theta_{D}} {32 \rho M c_{0}^{4}} \left(V_{0} + \frac {V_{1}} {3} \right),
\end{equation}
$M$ is a mass of atom in a lattice. Estimations show, that this parameter can be of the order of unity or less. Since it is supposed, that $\Lambda > 0$, then $\sigma > 1$ always holds. Transition to Debye theory occurs at $\Lambda = 0$ and $\sigma = 1$.\\
\indent At zero temperature renormalization of speed is determined by the equation 
\begin{equation}\label{sigma_0}
(\sigma_{0}^{2} - 1) \sigma_{0} = \Lambda.
\end{equation}
At low temperatures $\tau / \sigma_{0} \ll 1$, and taking into account (\ref{debye_small_x}, \ref{Phi_def}, \ref{Phi_approx}) one obtains
\begin{equation}\label{sigma_low}
\sigma \approx \sigma_{0} + \Lambda \frac {8 \pi^{4}} {15 (3\sigma_{0}^{2} - 1)} \left(\frac {\tau} {\sigma_{0}} \right)^{4}.
\end{equation}
At high temperatures $\tau / \sigma_{0} \gg 1$ equation (\ref{main_dimless}) reduces itself to biquadratic equation
\begin{equation}\label{sigma_biq}
(\sigma^{2} - 1) \sigma^{2} = \frac {8} {3} \Lambda \tau,
\end{equation}
solution of which is 
\begin{equation}\label{sigma_high}
\sigma = \frac {1} {\sqrt{2}} \sqrt{1 + \sqrt{1 + \frac {32} {3} \Lambda \tau}}.
\end{equation}
Here two typical temperature intervals can be separated. In the most probable case of small phonon--phonon interaction constant, even at high temperatures condition $8 \Lambda \tau / 3 \ll 1$ can be true. Then linear dependency appears
\begin{equation}\label{sigma_high_lin}
\sigma \approx 1 + \frac {4} {3} \Lambda \tau.
\end{equation}
At sufficiently strong phonon--phonon interaction, when $8 \Lambda \tau / 3 \gg 1$, one obtains
\begin{equation}\label{sigma_high_f}
\sigma \approx \left ( \frac {8} {3} \Lambda \right )^{1/4} \tau^{1/4}.
\end{equation}
\begin{figure}
  \includegraphics[width=\columnwidth]{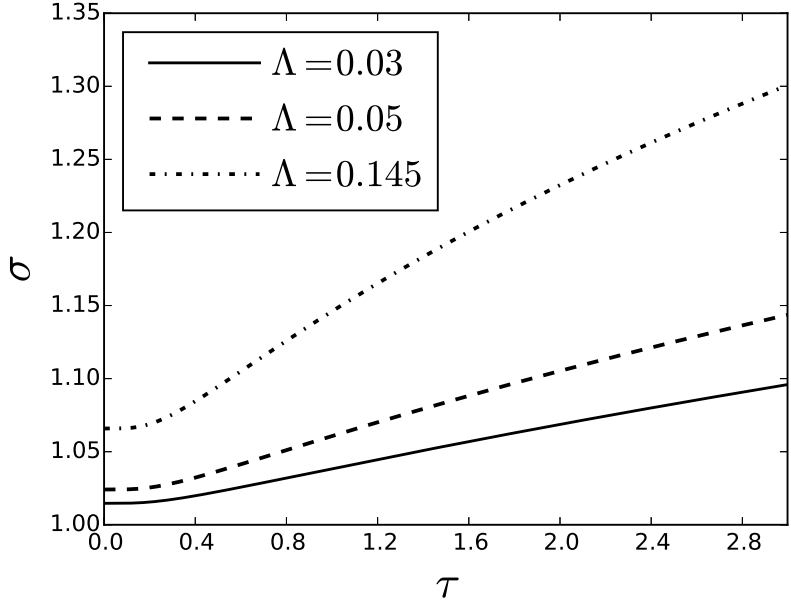}%
  \caption{\label{fig:sigma_tau}
  Parameter $\sigma \equiv \frac {c_{S}} {c_{0}} = \frac {\tilde \Theta_{D}} {\Theta_{D}}$ dependency on dimensionless temperature $\tau = \frac {T} {\Theta_{D}}$ for different values of parameter $\Lambda$.}
\end{figure}
Dependencies of parameter $\sigma$ on temperature for several values of $\Lambda$ are presented in figure \ref{fig:sigma_tau}.\\
\section{Thermodynamics of self-consistent phonon gas}
\indent Free energy (\ref{free_en}) can be expressed in the form
\begin{equation}\label{free_en_dimless}
\begin{split}
\frac {F} {N \Theta_{D}} = \frac {9} {16} \frac{(1 - \sigma^{2})}{\sigma} \Phi \left(\frac {\sigma} {\tau} \right) + \frac {9} {32} \frac {\Lambda} {\sigma^{2}} \Phi^{2} \left(\frac {\sigma} {\tau} \right) + \frac {9} {8} \sigma + \\ 
+ \tau \left[ 3 \ln \left(1 - e^{ - \frac{\sigma}{\tau}} \right) - D_{3} \left( \frac {\sigma} {\tau} \right) \right].
\end{split}
\end{equation}
If phonon interaction is neglected, $\Lambda = 0$ and $\sigma = 1$, then formula (\ref{free_en_dimless}) naturally turns into free energy of Debye theory \cite{Landau_Lifshitz}. Entropy is given by the equation
\begin{equation}\label{entropy}
S = - N \left[ 3 \ln \left( 1 - e^{ - \frac {\sigma} {\tau} } \right) - 4 D_{3} \left( \frac {\sigma} {\tau} \right) \right].
\end{equation}
Note, that expression for entropy can be obtained both from general equation for boson gas entropy \cite{Landau_Lifshitz} and from thermodynamic relation $S = - (\partial F / \partial T)_{V}$. When differentiating, owing to $\partial F / \partial c_{S} = 0$ condition, parameter $\sigma$ in (\ref{free_en_dimless}) should not be differentiated. Self-consistent phonon gas energy $E = F + TS$ has the form
\begin{equation}\label{en_dimless}
\frac{E} {N \Theta_{D}} = \frac {9} {16} \frac{(1 - \sigma^{2})} {\sigma} \Phi \left(\frac {\sigma} {\tau} \right) + \frac {9} {32} \frac{\Lambda} {\sigma^{2}} \Phi^{2} \left( \frac {\sigma} {\tau } \right) + \frac {9} {8} \sigma + 3 \tau D_{3}\left( \frac {\sigma} {\tau } \right).
\end{equation}
Pressure is obtained from $p = - (\partial F / \partial V)_{T}$ relation:
\begin{equation}\label{pressure}
\begin{split}
p = \frac {N} {V} \Theta_{D} \left[ \frac {9} {8} \sigma + 3 \tau D_{3} \left( \frac {\sigma} {\tau} \right) - \frac {9} {32} \frac{(\sigma^{2} - 1)} {\sigma} \Phi \left( \frac {\sigma} {\tau} \right) \right]\Gamma_{G} + \\
+ \frac {N} {V} \Theta_{D} \frac {9} {32} \frac{(\sigma^{2} - 1)}{\sigma} \Phi \left( \frac{\sigma} {\tau} \right) \Gamma_{\Lambda}.
\end{split}
\end{equation}
Two parameters enter into the equation (\ref{pressure})
\begin{equation}\label{grun}
\Gamma_{G} = - \frac {\partial \ln \Theta_{D}} {\partial \ln V}, \quad \Gamma_{\Lambda} = - \frac {\partial \ln \Lambda} {\partial \ln V}.
\end{equation}
Here $\Gamma_{G}$ is Gr\"uneisen parameter and $\Gamma_{\Lambda}$ is a new dimensionless parameter describing contribution of phonon--phonon interaction effects to pressure. Since sound speed $c_{0}$ and elastic moduli are considered to be independent from temperature, coefficients (\ref{grun}) also does not depend on temperature. Phonon--phonon interaction influence on solid body thermal expansion needs a separate consideration.
\section{Specific heat}
Equation for specific heat $C_{V} = T (\partial S / \partial T)_{V}$ can be obtained from expression for entropy (\ref{entropy}):
\begin{equation}\label{spec_heat}
C_{V} = 3N \left [ 4 \frac {\tau} {\sigma} D_{3} \left( \frac {\sigma} {\tau} \right) - \frac {3} {e^{ \sigma/\tau } - 1} \right] \left( \frac {\sigma} {\tau } - \frac {d\sigma} {d\tau} \right).
\end{equation}
Derivative of parameter $\sigma$ with respect to temperature entering into equation (\ref{spec_heat}) can be found from equation (\ref{main_dimless}), so that
\begin{equation}\label{deriv}
\frac {\sigma} {\tau} - \frac {d\sigma} {d\tau} = \frac{\sigma} {\tau}\cdot \frac{(1 - 3 \sigma^{2})} {1 - 3 \sigma^{2} + \frac {8} {3} \frac{\Lambda} {\sigma} \left[ \frac {3} {e^{\sigma/ \tau} - 1} - 4 \frac {\tau} {\sigma} D_{3} \left( \frac{\sigma} {\tau} \right) \right]}.
\end{equation}
At $\sigma = 1$ equation (\ref{spec_heat}) naturally gives the result of Debye theory. It is known, that the law of corresponding states, which consists in the fact, that specific heat and other thermodynamic quantities are functions of dimensionless temperature $\tau = T / \Theta_{D}$, takes place in the Debye theory \cite{Landau_Lifshitz}. Taking into account of phonon--phonon interaction leads to violation of this law and every specific phonon system is additionally characterized by its dimensionless parameter $\Lambda$. Let us consider low and high temperature limiting cases. %
\begin{figure}
  \includegraphics[width=\columnwidth]{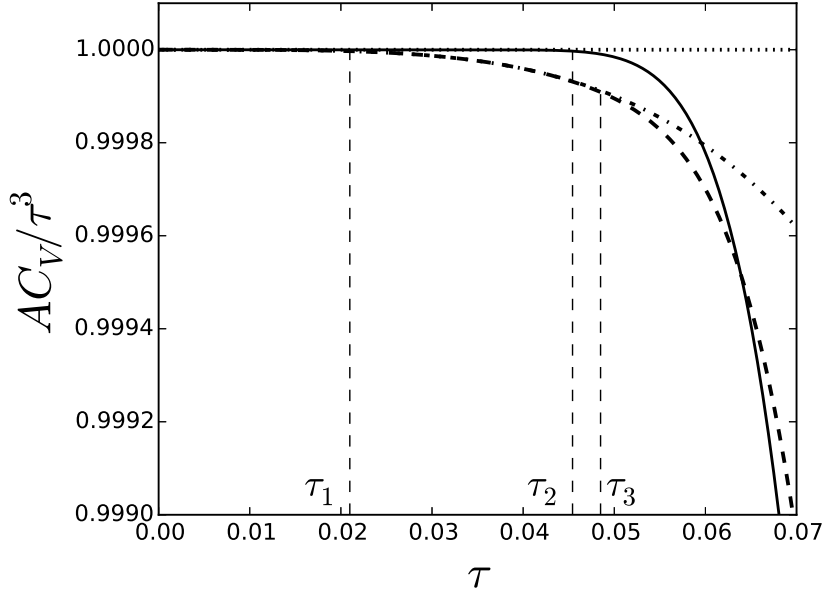}%
  \caption{\label{fig:spec_heat_norm}
    Specific heat behavior at low temperatures. Here $\Lambda = 0.145$ and $A = \frac {5 \sigma_{0}^{3}} {12 \pi^{4} N}$. Specific heats: solid line -- Debye theory; dashed line -- self-consistent phonons; dash and dot -- approximate equation (\ref{spec_heat_lt}). Typical temperatures: $\tau_{1} \approx 0.021$, $\tau_{2} \approx 0.045$, $\tau_{3} \approx 0.048$ (see explanations in the text).}
\end{figure}
\subsection{Low temperatures}
\indent At $T = 0$ parameter $\sigma = \sigma_{0}$ renormalizing sound speed is determined from equation (\ref{sigma_0}). At low temperatures $\sigma = \sigma_{0} + \sigma'$ and inequality $\sigma' / \sigma_{0} \ll 1$ is supposed, and from (\ref{sigma_low}) it is concluded, that
\begin{equation}\label{sigma_prime}
\sigma' = \Lambda \frac {8 \pi^{4}} {15 (3\sigma_{0}^{2} - 1)} \left(\frac {\tau} {\sigma_{0}} \right)^{4}.
\end{equation}
Then from (\ref{spec_heat}) for specific heat the following expression is obtained
\begin{equation}\label{spec_heat_lt}
C_{V} \approx \frac{12 \pi^{4}} {5} N \left( \frac{\tau} {\sigma_{0}}  \right)^{3} \left[1 - \Lambda \frac {56 \pi^{4}} {15 \sigma_{0} (3 \sigma_{0}^{2} - 1)} \left( \frac {\tau} {\sigma_{0}} \right)^{4} \right].
\end{equation}
In comparison with Debye theory specific heat behavior changed in two aspects. First, since $\sigma_{0} > 1$, coefficient at cubic dependence decreased, and second, there appeared a term proportional to $\tau^{7}$ besides the cubic one. \\
\indent Often Debye energy is determined from specific heat behavior at low temperatures \cite{Ashcroft_Mermin}, which according to (\ref{spec_heat_lt}) has the form
\begin{equation}\label{Cv_3}
\frac {C_{V}} {N} \approx \frac {12 \pi^{4}} {5} \left(\frac {T} {\tilde \Theta_{D0}} \right)^{3}.
\end{equation}
Here $\tilde \Theta_{D0} = \sigma_{0} \Theta_{D}$. As it can be seen, at low temperature specific heat measurements self-consistent Debye energy $\tilde \Theta_{D0}$ is measured, which differs from self-consistent energy $\tilde \Theta_{D} = \sigma \Theta_{D}$ at finite temperatures.\\
\indent In order for specific heat to satisfy $T^{3}$ law, the second term in (\ref{spec_heat_lt}) needs to be much less than the first one, i.e. a condition 
\begin{equation}\label{t3law_cond}
\left( \frac {\tau} {\sigma_{0}} \right)^{4} \ll \frac {15 \sigma_{0} (3 \sigma_{0}^{2} - 1)} {56 \pi^{4} \Lambda}
\end{equation}
should be satisfied. If $\Lambda$ is small, then $\sigma_{0} \approx 1$ and $(T / \Theta_{D})^{4} \ll 15 / 28 \pi^{4} \Lambda$. By means of measuring divergence of specific heat from cubic law at low temperatures, dimensionless constant $\Lambda$ can be determined. It should be noted, that there is no power correction term for specific heat at low temperatures in usual Debye theory and next order correction is of exponential character
\begin{equation}\label{Cv_D}
C_{V}^{D} \approx 3 N \left[ \frac {4 \pi^{4}} {5} \tau^{3} - \frac {3} {\tau} e^{- \frac {1} {\tau}} \right].
\end{equation}
\indent As it can be seen from (\ref{t3law_cond}), with increase of phonon--phonon interaction temperature domain of $\tau^{3}$ law applicability gets narrower. Thus, at $\Lambda = 0.1$ $\tau^{3}$ law is well satisfied at $\tau^{4} \ll 0.08$, and in case of $\Lambda = 1$ -- at $\tau^{4} \ll 0.05$. It is known, that for specific heat behavior to be close enough to $\tau^{3}$ law usually temperature range below $\tau \approx 1/50$ is needed \cite{Kittel}. Presence of power correction to low temperature specific heat proportional to $\tau^{7}$ and dependence of temperature domain of $\tau^{3}$ law applicability on phonon--phonon interaction constant explains, probably, the fact that $\tau^{3}$ law is observed at lower temperatures, than it could be expected. Temperature dependencies of specific heat ratio to temperature cube in Debye theory and in the considered approach are presented in figure \ref{fig:spec_heat_norm}. As it is seen from the plot, specific heat behavior divergence from cubic law begins significantly sooner for self-consistent phonons (at $\tau_{1} \approx 0.021$), than for Debye phonons (at $\tau_{2} \approx 0.045$). Temperature $\tau_{3} \approx 0.048$ shows, when values calculated by approximate formula (\ref{spec_heat_lt}) start to diverge from those calculated from (\ref{spec_heat}) equation. As a criterion of divergence of two curves $f_{1}(\tau)$ and $f_{2}(\tau)$ condition $2 \frac {|f_{1}(\tau) - f_{2}(\tau)|} {f_{1}(\tau) + f_{2}(\tau)} = 3 \cdot 10^{-6}$ was chosen. Discovery of correction proportional to $T^{7}$ in measurements of low temperature specific heat could verify validity of the proposed model.
\begin{figure}
  \includegraphics[width=\columnwidth]{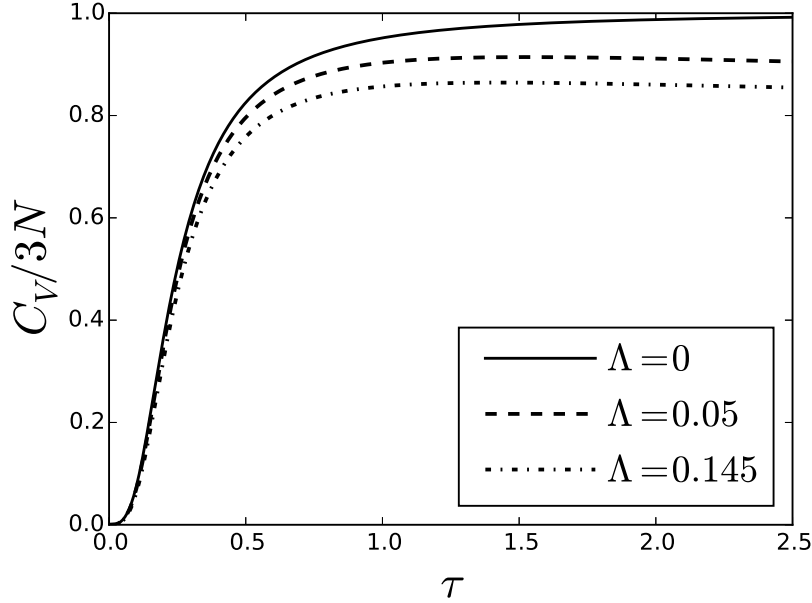}%
  \caption{\label{fig:spec_heat_full}
    Dependency of specific heat on temperature for different values of phonon--phonon interaction constant.}
\end{figure}
\subsection{High temperatures}
Let us consider case of high temperatures, so that $\tau \gg \sigma$ and equation (\ref{sigma_high}) is true. Here two cases are possible. At small $\Lambda$, besides $\tau \gg 1$ condition, $8 \Lambda \tau / 3 \ll 1$ can be satisfied, so that constraints on temperature are $\sigma \ll \tau \ll 3/8\Lambda$. Then in this range of temperature $\sigma \approx 1 + \frac {4} {3} \Lambda \tau$ (\ref{sigma_high_lin}) and linear on temperature deviation from Dulong--Petit law takes place
\begin{equation}\label{dulong_petit_dev}
C_{V} \approx 3 N \left(1 - \frac{4} {3} \Lambda \tau \right).
\end{equation}
Analogous dependency of the form $C_{V} = 3 N (1 - BT)$ was obtained in terms of absolutely different approach in \cite{Bazarov}. According to \cite{Bazarov} $B = 0.63 \frac {a^{6}} {z \varepsilon_{LJ} \sigma_{LJ}^{6}}$, where $\varepsilon_{LJ}$, $\sigma_{LJ}$ are parameters of Lennard--Jones potential, $a$ is a distance between the closest neighbors, $z$ is a number of the closest neighbors. Calculated using this formula values of dimensionless constant of phonon--phonon interaction for neon, argon and krypton are given in table \ref{table}. For all elements $z = 12$ and it was supposed, that $a / \sigma_{LJ} = 1.1$ \cite{Bazarov}. Calculated value of $B$ is in agreement with known experimental results for krypton \cite{Clusius}. Note that, however, ratio $a / \sigma_{LJ}$ should rise with temperature \cite{Bazarov}, therefore values of $B$ and $\Lambda$ in table \ref{table} are most likely somewhat understated.\\
\indent If both conditions $\tau \gg 1$ and $8 \Lambda \tau / 3 \gg 1$ are satisfied, which is possible at temperatures below the temperature of melting for solids with rather high phonon--phonon interaction, relation (\ref{sigma_high_f}) takes place and specific heat asymptotically approaches constant value
\begin{equation}\label{Cv_3_4}
C_{V} \approx \frac {3} {4} \cdot 3 N,
\end{equation}
which is $3/4$ of Dulong--Petit law value.\\
\indent The dependencies of specific heat on temperature for several values of $\Lambda$ are shown in figure \ref{fig:spec_heat_full}.
\begin{table}
\begin{center}
  \caption{Dimensionless parameter of phonon--phonon interaction}
    \begin{tabular}{| c | c | c | c |}
    \hline
      & Ne & Ar & Kr \\ \hline
    $\varepsilon_{KJ},\; K$ & 36 & 121 & 173 \\ \hline
    $\Theta_{D},\; K$ & 75 & 92 & 72 \\ \hline
    $B,\; K^{-1}$ & 0.0026 & 0.0008 & 0.0005 \\ \hline
    $\Lambda$ & 0.145 & 0.05 & 0.03 \\ \hline
    \end{tabular}
    \label{table}
\end{center}
\end{table}
\section{Self-consistent phonon interaction}
\indent Approximate accounting for phonon--phonon interaction in self-consistent field method with Hamiltonian (\ref{H_s}) has lead to renormalization of the phonon speed and appearance of its dependence on temperature. Unaccounted in this approximation interaction, which is contained in correlation Hamiltonian (\ref{H_c}), describes interaction of self-consistent phonons. Although the present work does not investigate effects connected with such interaction, let us give correlation Hamiltonian (\ref{H_c}) in terms of self-consistent phonon creation and annihilation operators
\begin{equation}\label{H_c_decomp}
H_{C} \equiv H_{ph-ph} = H_{C}^{(2)} + H_{C}^{(3)} + H_{C}^{(4)},
\end{equation}
where
\begin{equation}\label{H_c_2}
H_{C}^{(2)} = \frac {\hbar} {4 \rho c_{S}} \left( \lambda_{aibj} - \lambda \delta_{ab} \delta_{ij} \right) \sum_{\mathbf{k},\alpha ,\beta}\frac {k_{i} k_{j}} {k} e_{a}^{*} (k, \alpha) e_{b} (k, \beta) \left[ \psi_{\mathbf{k}\alpha }^{\dagger} \psi_{\mathbf{k} \beta} - \left\langle \psi_{\mathbf{k} \alpha}^{\dagger} \psi_{\mathbf{k} \beta} \right\rangle \right],
\end{equation}
\begin{equation}\label{H_c_3}
\begin{split}
H_{C}^{(3)} = \frac {1} {2 \sqrt{V}} \left( \frac {\hbar} {2 \rho c_{S}} \right)^{\frac {3} {2} } \left( \lambda_{aikj} \delta_{bc} + \frac{1} {3} \lambda_{aibjck} \right)\times \\ 
\times \sum_{\substack{\mathbf{k}_{1}, \mathbf{k}_{2}, \mathbf{k}_{3} \\ \alpha, \beta, \gamma}} \frac{ k_{1i} k_{2j} k_{3k}} {\sqrt{ k_{1} k_{2} k_{3}}}\Delta \left( \mathbf{k}_{1} + \mathbf{k}_{2} + \mathbf{k}_{3} \right) e_{a} (k_{1}, \alpha) e_{b} (k_{2}, \beta) e_{c} (k_{3}, \gamma) \psi_{\mathbf{k}_{1} \alpha} \psi_{\mathbf{k}_{2} \beta } \psi_{\mathbf{k}_{3} \gamma}, 
\end{split}
\end{equation}
\begin{equation}\label{H_c_4}
\begin{split}
H_{C}^{(4)} = \frac{ \hbar^{2}} {16 \rho^{2} \tilde c_{S}^{2} V} \left(\frac {1} {2} \lambda_{ijkl} \delta_{ab} \delta_{cd} + \lambda_{aibjkl} \delta_{cd} + \frac {1} {6} \lambda_{aibjckdl} \right) \times \\ 
\times \sum_{\substack{ \mathbf{k}_{1}, \mathbf{k}_{2}, \mathbf{k}_{3},\mathbf{k}_{4} \\ \alpha ,\beta ,\gamma ,\delta}} \frac{ k_{1i} k_{2j} k_{3k} k_{4l}} {\sqrt{k_{1} k_{2} k_{3} k_{4}}} \Delta \left(\mathbf{k}_{1} + \mathbf{k}_{2} + \mathbf{k}_{3} + \mathbf{k}_{4} \right) e_{a}( k_{1},\alpha) e_{b} (k_{2},\beta) e_{c} (k_{3},\gamma) e_{d} (k_{4},\delta)\times \\ 
\times \left[ \psi_{\mathbf{k}_{1} \alpha} \psi_{\mathbf{k}_{2} \beta}\psi_{\mathbf{k}_{3} \gamma} \psi_{\mathbf{k}_{4} \delta} - \left\langle \psi_{\mathbf{k}_{1} \alpha} \psi_{\mathbf{k}_{2} \beta}\psi_{\mathbf{k}_{3} \gamma} \psi_{\mathbf{k}_{4} \delta} \right\rangle  \right]. 
\end{split}
\end{equation}
Here $\psi_{\mathbf{k} \alpha} = \psi_{- \mathbf{k} \alpha}^{\dagger} = b_{\mathbf{k} \alpha} + b_{- \mathbf{k} \alpha}^{\dagger}$ (\ref{chi_psi}), also $\Delta(\mathbf{k}) = 1$ if $\mathbf{k} = 0$ and $\Delta(\mathbf{k}) = 0$ if $\mathbf{k} \ne 0$. Contribution of interaction between self-consistent phonons (\ref{H_c_decomp}) to thermodynamics of the system can be taken into account by means of perturbation theory and using diagram technique \cite{Poluektov3}. Kinetics of self-consistent phonon gas can be built based on interaction Hamiltonian (\ref{H_c_decomp}) \cite{Zaiman_2,Gurievich}. 
\section{Conclusions}
\indent Self-consistent field approach is used to describe interacting phonons in a solid body. Accounting for phonon--phonon interaction in self-consistent field approximation leads to appearance of phonon speed dependence on temperature. Introduced in this approach phonons can be called "self-consistent" or "dressed" in contrast to "bare" phonons in Debye theory. Thermodynamics of "self-consistent" phonons system is constructed. Phonon specific heat is calculated and it is shown, that at low temperatures there is a power correction to $T^{3}$ law proportional to $T^{7}$. This, apparently, allows to explain why $T^{3}$ law is only observed at rather low temperatures. Observation of $T^{7}$ dependence, alongside with cubic one, in an experiment could be a correctness criterion for the proposed theory. At high temperatures deviation from Dulong--Petit law proportional to $T$ is predicted, which is observed in experiment. \\
\indent As far as the present work introduces notion of different types of quasiparticles, namely "bare" (or Debye) phonons and "dressed" (or self-consistent), so in conclusion general remarks are made about quasiparticle description of many--body systems. Notion of quasiparticles is an approximation by it's nature, since if there was a possibility to solve exactly many--body problem, there would be no need to introduce a notion of quasiparticles. However, since exact solution of many--body problem can not be found, notion of quasiparticles is, undoubtedly, fundamental, although introduction of quasiparticles itself for one or another system is ambiguous. A question 'Which from many possible definitions of quasiparticles is the most "real"?' can be asked. Obviously, the most "real" of all quasiparticles should be considered those ones, which in the best way describe a given system in the ideal gas model of such quasiparticles. Self-consistent field model is constructed in such way that quadratic Hamiltonian of the model is chosen to be the closest to the exact Hamiltonian of the described system \cite{Poluektov1,Poluektov2,Poluektov3}, and consequently quasiparticles emerging within the scope of such model are the best to describe the given system. For further improvement of description, interaction between quasiparticles should be taken into account. This remark fully applies to "self-consistent" phonons, with the use of which, as shown above, it is possible to describe more subtle effects than with phonons in Debye approximation.

%
%

\end{document}